\documentclass{amsart}
\usepackage{graphicx}
\theoremstyle{definition}

\theoremstyle{remark}

\numberwithin{equation}{section}
\begin{document}

\title[SOLITON ORGANIZATION of THERMAL FIELD...]{SOLITON ORGANIZATION of THERMAL FIELD in
a CHAIN at HIGH TEMPERATURE}%
\author{Leonid S. Metlov}%
\address{Donetsk Institute of Physics and Engineering, 72, R. Luxembourg str.,
83114 Donetsk, Ukraine}%
\email{metlov@atlasua.net}%

\thanks{    This work was supported by the Fond of Fundamental Researches of
the National Academy of Science of Ukraine (Grant No 4.4/242) and
by Donetsk
Innovation Center (provision of computer and interest to the work).}%

%\date{05.03.2001}%
%\dedicatory{}%
%\commby{}%
% ----------------------------------------------------------------
\begin{abstract}
Thermal field soliton self-organization arising due to absorption
of background  atoms vibrations is observed in numerical
experiment in nonlinear chain with Lennard-Jones potential at high
temperature. At some stage intensive space-localized waves are
formed and give additional peaks on high-energy tile of energy
distribution unlike of Gibbs one.
\end{abstract}
\maketitle
% ----------------------------------------------------------------
\section*{Introduction}

A phonon (the extended mode) delocalized in the coordinate space,
but localized in the wave space is the fundamental quasiparticle
in crystal reality. Since any of a number of phonons occupies the
entire space of the crystal, each individual particle takes part
in the motion of all the phonons. In the result of accidental
phase combination of some phonon groups the multiphase and
single-phase waves may arise with spatial localization of the
order of the lattice spacing. In consequence of disperse features
of the lattice they will be scattered prior to their wave
properties will be exhibited. With nonlinearity phonon groups may
form a long-lived bound states and in this situation the wave
properties will be exhibited in full measure. The soliton-like
excitations of two types, such as kinks and bell-form solitons
 -~\cite{CBG94,PSZ96,SHP79,ZOW95,XZ96}, on the one hand, or breasers (gap
solitons, intrinsic localized modes and so on)
~\cite{CBG94,DP93,DPB93,GM95,Kiv93}, on the other hand, observed
in computer experiments for different nonlinear lattices, directly
confirm this fact. The former may exist below the linear phonon
band, the latter ones above the upper harmonic phonon band edge,
that is inside the gap ~\cite{GM95} or inside the self-induced gap
~\cite{Kiv93}.

Nonlinear waves of different nature manifest a series of common
properties including suppression and absorption of small waves by
larger ones. In Ref.~\cite{DP93} such phenomenon was observed for
small and large exact breasers. In Ref. ~\cite{CDRT98} the chaotic
breasers were observed in the numerical experiment for the
FPU-chains, which manifest the effect of energy concentration by
separate excitations. As the initial condition the  $\pi$-mode was
taken. It is shown that the $\pi$-mode decays into local
excitations with chaotic behavior. In the initial stage the
concentration of energy by separate  breasers grows. Latter on
these excitations vanish and the system approaches the
equipartition state. The authors state that the "pumping" of
energy from the high-frequency $\pi$-mode to the low-frequency
region of phonon spectrum takes place by this way.

After relaxation the chaotic breasers die out because of
"starvation". This brings up the question: is such behavior of a
system, when the chaotic breasers do not die out and rest after
the relaxation, possible? To achieve such regime it is necessary,
that the mean energy per one particle, that is, temperature, be
high enough. The localization modes are regarded also for
FPU-chains in numerical experiment of Ref. ~\cite{FHLZ98}.

The aim of this paper is to investigate similar effects for kink
solitons at high temperature and their influence on thermal field
by means of computer simulation. To minimize the non-equilibrium
pumping effect here the "white noise" distribution is used as the
initial state. The "white noise" is energy distribution closest to
thermodynamic equilibrium state. One may expect, that the
relaxation passes fast enough.

\section{Basic Statements}

One may note that a FPU-chain is a very idealistic model for
atomic systems and it deviates from realistic ones to a greater
extend with the vibration amplitude increase. In our case such a
situation presents the main interest for the investigation. It
follows herefrom that one of the realistic potentials containing
non-linearity of all orders must be used for such system at high
temperature.

All simulation was performed for a chain built up from 200
particles interacting via the most popular 6-12 Lennard-Jones'
potential, that is:

\begin{equation}\label{e:a1}
  U_{ij}=u_{min}\{(r_{0}/r_{ij})^{12}-2(r_{0}/r_{ij})^{6}\}
\end{equation}

where $r_{ij}=\sqrt{(X_{i}-X_{j})^{2}+(Y_{i}-Y_{j})^{2}}$- is the
distance between particles with numbers $i$ and $j$, and Cartesian
coordinates $X_{i}, Y_{i}, X_{j}, Y_{j},$ $r_{0}$ - - is the
equilibrium distance for two-particle systems, $u_{min}$ - is the
binding energy. For the sake of simplicity the reduced units of
length and time as $\widetilde{r_{ij}}=r_{ij}/r_{0}$ and
$\widetilde{t}=t/T$, respectively, are defined. Here
$T=2\pi\sqrt{m/U''_{ij}(0)}$ is the period of small vibrations,
$m$ is the particle mass which equals unity thereafter. The
binding energy in the reduced units turned out to be equal to
$2(\pi/6)^{2}$. A smaller than $T$ conventional unit of time must
be introduced for the description of wave phase. We take it to be
equal to $t_{con}=0.005$ (time step in the computer experiment).
Sixth-order Yoshida's symplectic algorithm is used for calculation
~\cite{Yo90}. The law of conservation of energy is fulfilled to
relative accuracy $10^{-6}$.

The equilibrium state is determined by coordinates $X^{(0)}_{i},
Y^{(0)}_{i}$:

\begin{equation}\label{e:a2}
X^{(0)}_{i}=r_{0}i,  Y^{(0)}_{i}=const
\end{equation}

    The initial distribution of particle velocities is chosen in the
form of white noise with amplitude 4.078 for X- polarization only.
The initial disturbances along Y-direction was not preset,
however, for additional control over computer errors the movement
in this direction was permitted. The average additional energy per
one particle contributed by white noise turned out to be equal to
2.6895 of reduced units. This value is 4.9 times as much as the
binding energy. Such correlation of energies corresponds to high
enough temperature of the chain which is much higher than the
melting point.

\section{Results and Discussions}

The evolution of white noise in time for a chain built up from 200
particles is presented by four graphic fragments. Every fragment
is composed of 46 constitutive spatial profiles reflecting the
square of particle velocity or kinetic energy versus its number in
the chain. In its turn, every constitutive profile is formed by
graphic superposition of ten elementary profiles taken during ten
time steps. Such display permits one to improve the tracing of
waves for a long time.

Really as would be expected, the early stage of evolution (the
upper fragment in fig.1) is very close to homogeneous
distribution. Nevertheless at this stage the local excitations
with size of the order of lattice spacing are traced already.
These local objects conserve their shape both at free spreading,
and after interactions in between for a long time. Their
velocities determined by slope of traces vary from 10 to 16 of
reduced units and they turn out to be proportional to wave
amplitude. A consequence of the above is the fact that we deal
with the soliton-like waves.

\begin{figure}[p]
  \includegraphics[width=6.5in]{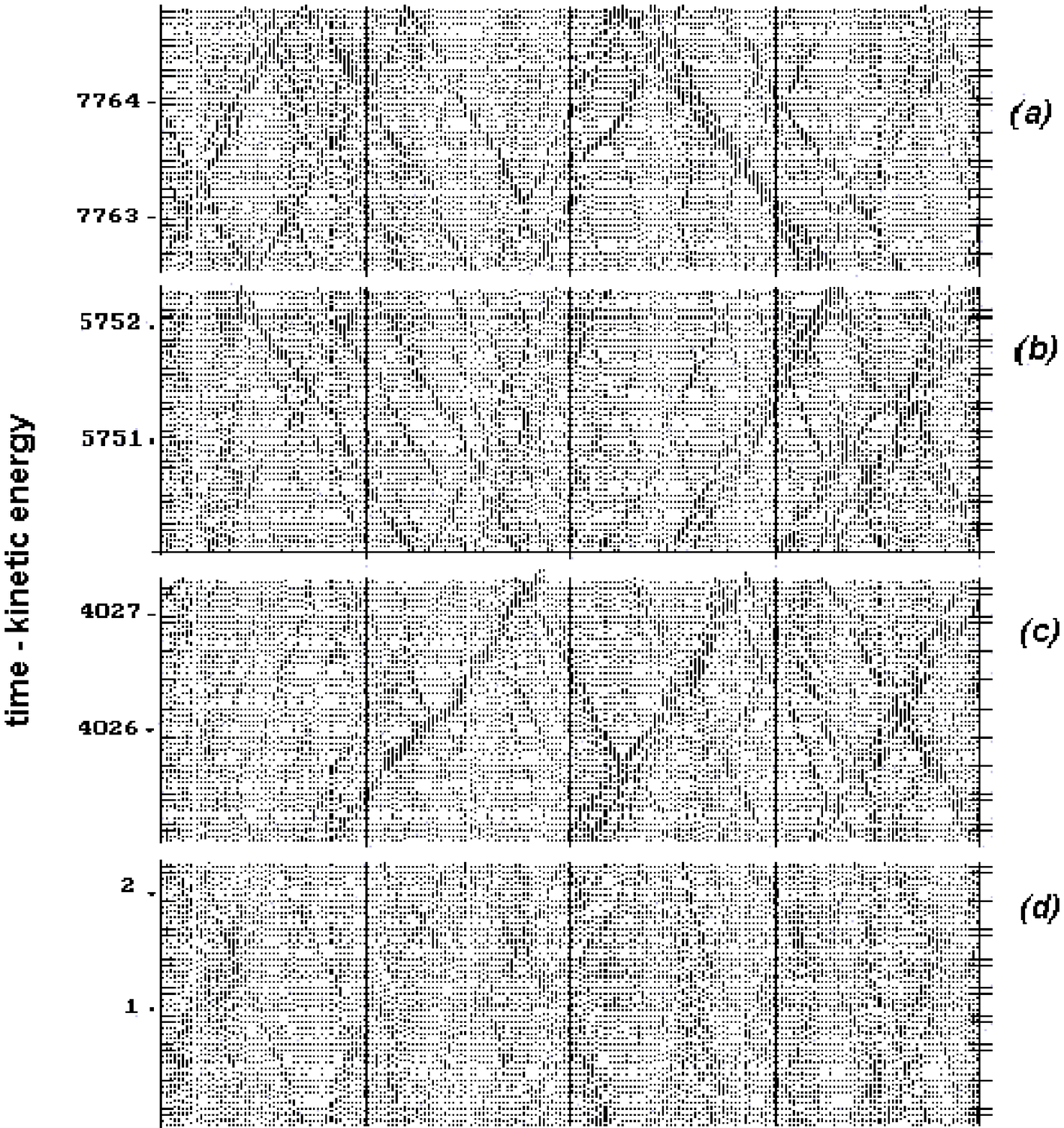}\\
  \caption{Evolution of white noise: Reading of time along vertical}\label{f1}
\end{figure}

\begin{figure}[p]
  \includegraphics[width=5in]{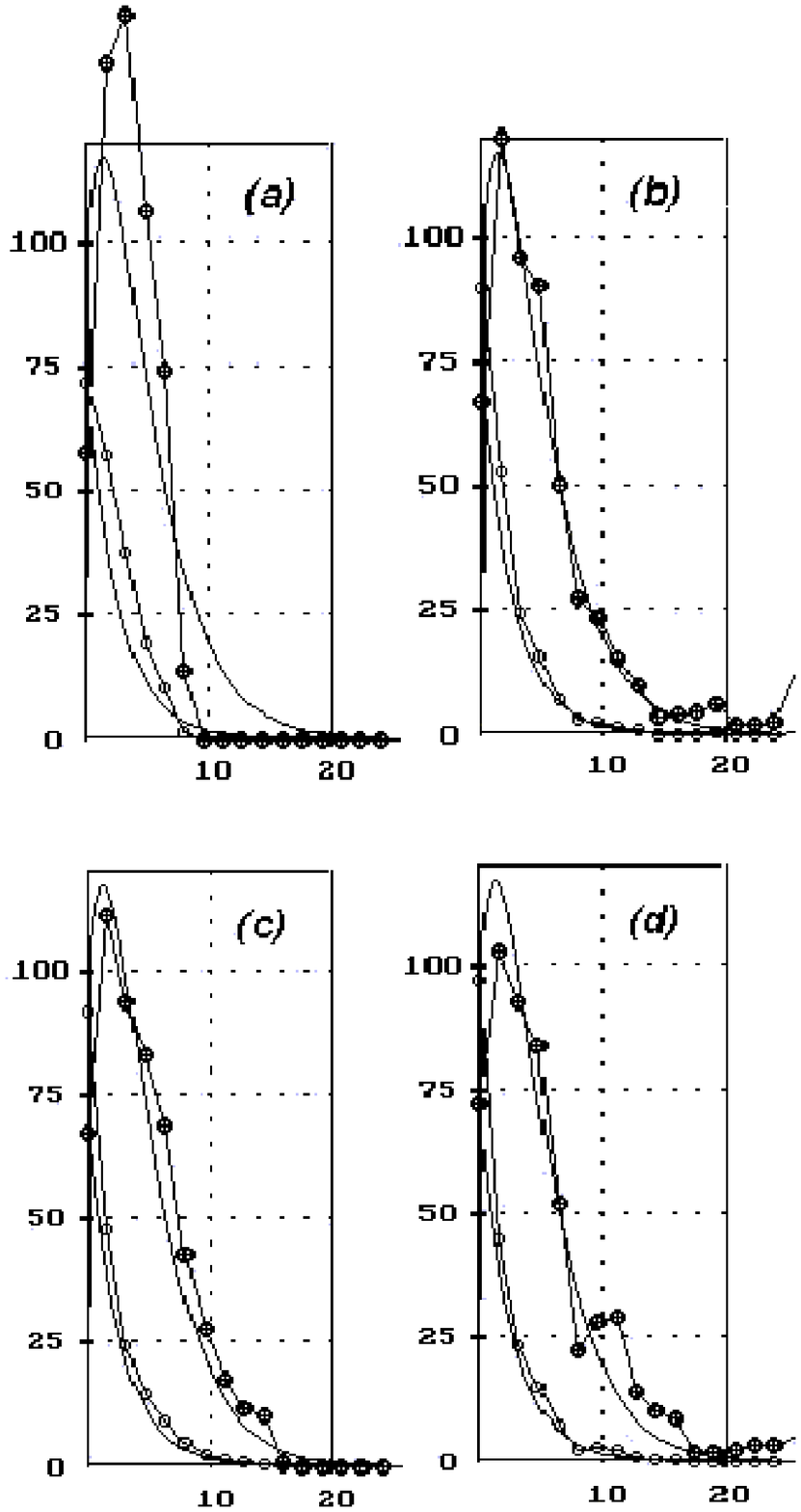}\\
  \caption{"Experimental" and theoretical distributions of particles and energy
  spectra versus its energy}\label{f2}
\end{figure}

From general considerations it is quite evident that they are the
soliton-like objects of kink kind. Really, if spreading of
excitation is presented as consecutive frontal collisions of
particles, then in the result of a separate collision a moving
particle stops and the resting particle starts to move with the
same velocity. In the next collision this moving particle stops
too in the new position, and so on. As a result, we have the
switching wave of the kink type. Evidently, in our case we have a
multitude of such waves, interacting in addition each with other.

By virtue of white noise proximity to the equilibrium state if may
seem that the relaxation will pass sufficiently fast. However, the
next, second fragment on the top of fig. 1 testifies that the
system approaches equipartition state, but deviated towards even
greater inhomogeneity of the spatial distribution of particle
velocities. In the chain the high-amplitude waves arise,
concentrating a significant part of system energy. As the law of
energy conservation is fulfilled with high precision, this can be
traced at the cost of energy re-distribution among the
excitations, namely, high excitations have captured the energy of
the weak ones. This phenomenon is analogous to the decay of the
For calculation the $\pi$-mode into the chaotic breasers
~\cite{CDRT98}. The surprising thing is that almost identical
effect accompanies the breakdown of white noise. Next two
fragments (fig. 1) show, that the level of inhomogeneity of the
velocity field varies in time, however in any event it is
considerably above than at the starting stage (the uppermost
fragment). This inhomogeneity don't vanish at high time of system
evolution and, it is obvious that it don't vanish at all. Thus,
while in the Ref.~\cite{CDRT98} chaotic breasers were an interface
at system relaxation and they vanished after the achievement of
thermal equilibrium, in our case the soliton-like excitations are
the main objects of the equilibrium thermal field. One may
conclude that solition organization of thermal field takes place
in such situation.

To support this conclusion let us correlate calculated
distribution of particle energy with theoretical Gibbs one. In
fig. 2 the curves with circles correspond to "real" data of the
numerical experiments, solid ones correspond to theoretical
distribution: Descending curves describe a distribution of
particles by energy  , determined as number of particles with
energy
$(i-1)\triangle\varepsilon\leq\varepsilon<i\triangle\varepsilon$,
where $i$ - is the number of interval. The curves with one or many
peaks are the energy spectra , determined as total energy of
particles in the same interval. Theoretical equilibrium
distribution of this quantities are of the form:

\begin{eqnarray}\label{e:a3}
\nonumber
  n_{i}=n_{0}\exp(-\varepsilon_{i}/\theta), \\
  E_{i}=\varepsilon_{i}n_{i}
\end{eqnarray}

with the fulfillment of laws of conservation for particles and
energy:

\begin{eqnarray}\label{e:a4}
\nonumber
  N=\sum_{i=1}^{\infty}n_{i}, \\
  E=\sum_{i=1}^{\infty}\varepsilon_{i}n_{i}
\end{eqnarray}

where $n_{0}$ - is the number of particles with minimum energy,
$\theta$ - is temperature in the energy representation of reduced
units.

From the last relations the expressions for $n_{0}$ and $\theta$
follow:

\begin{eqnarray}\label{e:a5}
\nonumber
 \theta=E/N , \\
 n_{0}=N^{2}/E
\end{eqnarray}

All graphs built with numerical experiments are obtained by means
of averaging over 460 time steps and they correspond to fragments
in fig 1. Energy interval $\triangle\varepsilon$ used at building
of graphics was taken to be equal to 1.6 of reduced units.

One can see from the first graph that the white noise deviates
from Gibbs distribution to a greater extend than any in fig. 2. On
the energy spectra for real data peak has higher value than for
theoretical ones at the cost of a fraction of low-energy
excitations. The particles with high energy are absent in the
system at all. With the evolution of white noise the energy is
"pumped" from the low-energy excitation to the high-energy ones.
One can see from the second graph in fig. 2 that the first peak
has been decreased below theoretical value while additional peaks
have arisen in the high-energy region. In this region the energy
curve lies markedly higher than the theoretical one. With latter
evolution (the third graph in fig. 2) a peculiar kind of returning
is observed, namely, real distribution achieves the theoretical
one. Additional peaks have vanished. It should be noted that such
equilibrium differs in the nature of its organization from a
phonon variant ~\cite{CDRT98}. Really, in our case (see the third
fragment in fig. 1) local excitations continue to exist during all
the time. This is the principal new finding.

One may expect that the system having achieved, finally, the
thermal equilibrium, will rest in this state for a long time.
However, the further evolution of the system (see the fourth graph
in fig. 2) shows that with time the system returns to the state
with additional peaks again. At the same time the returning can't
be considered as simple fluctuation, since it develops directly
and sequentially during a long time of about 300-500 reduced units
of time. Latter on both of these states alternate in time.

The departure of equilibrium distribution from the Gibbs one is
marked in Ref. ~\cite{SHP79} for chock waves in one-dimensional
chains. If $\theta_{0}$ is the initial temperature and $\theta$ is
the final temperature ahead and behind of the chock front,
respectively, then the equilibrium distribution of particles
velocities is determined by the following expression
~\cite{SHP79}:

\begin{equation}\label{e:a6}
  n_{i}=\frac{1}{2}(2k\theta)^{-\frac{1}{2}}[\exp(-\frac{(\nu_{i}-
  \vartheta^{\frac{1}{2}})^{2}}{2\theta_{0}})+\exp(-\frac{(\nu_{i}+
  \vartheta^{\frac{1}{2}})^{2}}{2\theta_{0}})]
\end{equation}

where $\vartheta=\theta-\theta_{0}$, $\nu_{i}$ - is velocity of a
particle, or with regard of potential energy contribution:

\begin{equation}\label{e:a7}
  n_{i}=\bar{n}_{0}\exp(-\frac{\varepsilon_{i}}{\theta_{0}})\coth(2\varepsilon_{i}\theta)^{\frac{1}{2}}
\end{equation}

where $\bar{n}_{0}=n_{0}\exp(-\frac{\vartheta}{2\theta_{0}})$ ,
$n_{0}=(2\pi\theta_{0})^{-\frac{1}{2}}$ - is the same $n_{0}$ as
in the expression ~(\ref{e:a3}).

One may see, that at $\vartheta>0$ $\bar{n}_{0}<n_{0}$ is true,
that is, the number of particles with minimum energy is less then
it for the Gibbs distribution. In the same time owing to factor
$\coth(2\varepsilon_{i}\theta)^{\frac{1}{2}}$ a share of the
high-energy particles grows. The same is true for energy spectrum.

From third graph in fig. 2 one may see, that the "experimental"
distribution being qualitatively described by the expression
~(\ref{e:a7}) is formed in the system. However no peaks in the
spectra in fig. 2b, d are described by this expression. The
emergence of the ones is quantitatively new peculiarity of a
non-linear system at high temperature. Its emergence obviously
connects with the long-living local high-energy excitations.

Thus the described computer experiment has shown that the
fundamental soliton organization of thermal wave field is possible
at high temperature. Such phenomenon has led to spontaneous
production of intensive nonlinear waves, which can be considered
as a peculiar kind of self-organization of a thermal field.
Investigation of this phenomenon is important to gain a better
understanding of the break of solid.

% ----------------------------------------------------------------
\bibliographystyle{amsplain}
\bibliography{forArXiv}

\end{document}